\documentclass[10pt,conference]{IEEEtran} 
\IEEEoverridecommandlockouts

\usepackage{listings}
\usepackage{xcolor}

\lstdefinelanguage{YAML}{
  sensitive=true,
  comment=[l]{\#},
  morestring=[b]',
  morestring=[b]",
  morecomment=[s]{/*}{*/},
  morekeywords={true,false,null,yes,no,on,off},
}

\lstdefinestyle{yamlstyle}{
  language=YAML,
  basicstyle=\ttfamily\small,
  backgroundcolor=\color{gray!6},
  frame=single,
  framerule=0.4pt,
  rulecolor=\color{black!30},
  numbers=left,
  numberstyle=\tiny\color{black!50},
  stepnumber=1,
  numbersep=8pt,
  showstringspaces=false,
  tabsize=2,
  breaklines=true,
  columns=fullflexible
}

\usepackage{booktabs}
\usepackage{tabularx}
\usepackage[flushleft]{threeparttable}
\usepackage{url}
\usepackage{cite}
\usepackage{amsmath,amssymb,amsfonts}
\usepackage{algorithmic}
\usepackage{graphicx}
\usepackage{textcomp}
\usepackage{xcolor}
\usepackage{float}
\usepackage{subcaption}
\newcommand{\linebreakand}{%
  \end{@IEEEauthorhalign}
  \hfill\mbox{}\par
  \mbox{}\hfill\begin{@IEEEauthorhalign}
}
\newcommand{\rqn}[1]{RQ\textsubscript{#1}}
\newcommand{\rp}{\footnote{\url{https://doi.org/10.5281/zenodo.17469262}}}

\usepackage{color}

\newcommand{\travis}{{\fontfamily{cmtt}\selectfont{.travis.yml }}}
\newcommand{\traviscommand}[1]{{\fontfamily{lmss}\selectfont{#1}}}
\newcommand{\tool}[1]{\texttt{#1}}

\usepackage[most]{tcolorbox}
\newtcolorbox{findingbox}{
  enhanced,
  colback=gray!10,   % light gray background
  colframe=black!80,  % subtle border
  boxrule=0.5pt,     % thin border
  arc=3pt,           % rounded corners
  left=8pt, right=8pt, top=6pt, bottom=6pt,
  before skip=6pt, after skip=0pt
}

\def\BibTeX{{\rm B\kern-.05em{\sc i\kern-.025em b}\kern-.08em
    T\kern-.1667em\lower.7ex\hbox{E}\kern-.125emX}}
\begin{document}

\title{
Investigating CI/CD-based Technical Debt Management in Open-source Projects
}

\author{
    \IEEEauthorblockN{João Paulo Biazotto}
    \IEEEauthorblockA{
        University of Groningen \\ 
        Groningen, The Netherlands \\
        University of São Paulo\\
        S\~ao Carlos - SP - Brazil \\
        j.p.biazotto@rug.nl
    }
    \and
    \IEEEauthorblockN{Daniel Feitosa}
    \IEEEauthorblockA{
        University of Groningen \\
        Groningen, The Netherlands \\
        d.feitosa@rug.nl
    }
    \and
    \IEEEauthorblockN{Paris Avgeriou}
    \IEEEauthorblockA{
        University of Groningen \\
        Groningen, The Netherlands \\
        p.avgeriou@rug.nl
    }
    \and
    \IEEEauthorblockN{Elisa Yumi Nakagawa}
    \IEEEauthorblockA{
        University of São Paulo\\
        São Carlos-SP, Brazil \\
        elisa@icmc.usp.br
    }
}

\maketitle

\begin{abstract}
Managing technical debt (TD) is critical to ensure the sustainability of long-term software projects. However, the time and cost involved in technical debt management (TDM) often discourage practitioners from performing this activity consistently. Continuous Integration and Continuous Delivery (CI/CD) pipelines offer an opportunity to support TDM by embedding automated practices directly into the development workflow. Despite this potential, it remains unclear how TDM tools could be integrated into CI/CD pipelines, and we still lack established best practices for this process. To address this problem, the objective of this study is to understand how TDM tools have been used in CI/CD pipelines and also identify potential configuration anti-patterns. To this end, we conducted a large-scale mining software repository (MSR) study on GitHub. In total, we collected around 600,000 Travis CI configuration files and 50,000 supporting scripts, and identified 3,684 pipelines that contain at least one TDM tool. We applied descriptive statistics to analyze the prevalence of tools and anti-patterns, and our findings show that most tools are executed and integrated using an external script; in addition, \textit{Absent Feedback} is the most common configuration anti-pattern. We believe that researchers and practitioners can use the evidence of this study to further investigate how to improve both the tools that are integrated in CI/CD and the integration practices.
\end{abstract}


\section{Introduction}
\label{sec:intro}
 
Over the years, several technical debt management (TDM) activities, such as identifying and measuring debt, have been developed to support developers and companies to keep technical debt (TD) under control~\cite{Li2015, Rios2018a}. Although paramount for the long-term sustainability of software projects, such TDM activities are often effort-intensive and require substantial time that could potentially be used to develop new features~\cite{Besker2018}. In this context, tooling is essential to automate or semi-automate such activities, and several recent studies investigated available tools~\cite{Silva2022}, practitioners' concerns about those tools~\cite{Biazotto2025b}, and their usage~\cite{Biazotto2023}. Although such studies list many available tools, they also highlight that the adoption of tools is a complex matter for many reasons, such as usability and workflow friction~\cite{Rios2020}, false positives~\cite{Sas2023}, and integration overheads \cite{Ochoa_2022}. These barriers help explain why teams underuse tools even when they acknowledge their benefits~\cite{Martini2018}.

To address some of those problems, the Manifesto on Reframing Technical Debt~\cite{Avgeriou2025} explicitly calls for tools that are easily integrated into existing workflows. Specifically, the manifesto suggests  to ``\textit{develop workflow-based TDM tools}'', emphasizing the need for solutions that integrate seamlessly into existing development processes. The same concern was reported by practitioners~\cite{Biazotto2025b}, who highlighted that TDM tools must be consistent with their workflows to reduce friction in tool adoption. 

One potential solution is to integrate TDM tools in Continuous Integration and Continuous Delivery (CI/CD) pipelines~\cite{Carvalho2024}. CI/CD aims to make the integration and delivery of changes safe, fast and routine~\cite{FITZGERALD2017176} and is widely adopted in open-source and industrial projects~\cite{Gallaba2020}. In addition, a previous systematic review showed that adopting CI/CD pipelines that contain TDM tools (e.g., static analyzers, code smell detectors, architectural conformance checkers) can identify TD earlier and reduce manual effort, turning TDM from sporadic activity into a continuous practice~\cite{Biazotto2023}.

Previous studies have reported that many existing TDM tools \textit{can} be integrated into CI/CD pipelines~\cite{Silva2022}. However, we lack consolidated evidence on the extent that TDM tools are \textit{actually} used in CI/CD pipelines. There is also little knowledge on \textit{how} TDM tools could be integrated in such pipelines, and \textit{when} the tools are executed within the pipelines (e.g., before or after deployment). To make matters worse, the pipelines themselves suffer from configuration problems, such as anti-patterns~\cite{Vassallo2019,Gallaba2020}, which degrade feedback quality, delay TD reports, and reduce developer trust. This suggests that simply adding a TDM tool to a CI/CD pipeline is insufficient; \textit{how} the tool is wired and governed matters. In summary, \textbf{the main problem we tackle in this study is the lack of large-scale empirical knowledge on how and when TDM tools are actually integrated into CI/CD pipelines, and what anti-patterns may occur and impede the pipelines. This can hinder the adoption of such tools and lead to less effective TDM using CI/CD pipelines.}

To address this problem and provide support to developers who aim at adopting TDM tools in CI/CD pipelines more systematically, we carried out a large-scale mining software repository (MSR) study on GitHub repositories. Specifically, we analyzed around \textbf{600,000} CI/CD configuration files and \textbf{50,000} supporting scripts. We characterized the pipelines to understand how TDM tools are integrated, when this happens in the pipeline, and the prevalence of configuration anti-patterns. Our study provides the following contributions:

\begin{itemize}
    \item \textbf{We systematically identify how TDM tools are executed within pipelines} (e.g., direct calls in the config vs. invocation through external scripts), revealing common integration patterns that could support developers planning the integration of tools.

    \item \textbf{We determine the specific pipeline stages, jobs, and phases} in which TDM tools are typically run, providing a broader view of  the tools within the CI/CD workflows\footnote{In this study, pipeline(s) and workflow(s) are used as synonyms.}.

    \item \textbf{We quantify how often key CI/CD anti-patterns} (e.g., skipped failures) occur in configurations that involve TDM tools, offering actionable evidence for practitioners and tool vendors.
\end{itemize}

The remainder of this paper is organized as follows: Section~\ref{sec:rw} discusses the related work and compares it to our results, highlighting our contributions.  Section~\ref{sec:sd} describes the methods used in this study, including data source selection, data collection pipeline, and data analysis approaches. Section~\ref{sec:results} presents the results, while  Section~\ref{sec:discussion} discusses them and points out implications for researchers and practitioners.
Section~\ref{sec:tov} outlines the threats to the validity of this study and the actions taken to mitigate them. Finally, Section~\ref{sec:conclusion} presents the conclusions of this study and discusses potential future research directions.

\section{Background and Related Work}
\label{sec:rw}

\subsection{Tooling for Technical Debt Management}
\label{sec:rw-tdm-tools}

Tooling is central to contemporary engineering practices, including continuous software development, agile methods, and DevOps, and supports nearly every phase of the life cycle, from design to deployment and maintenance~\cite{Theunissen2021}. Such tools encompass a wide range of artifacts and services, such as integrated development environments (IDEs), project and issue trackers, static analyzers, and testing frameworks; collectively, these tools enable teams to create, analyze, and manage software products \cite{Theunissen2021}.

Despite the broad adoption of general-purpose tools in daily workflows, \textbf{TDM tools} remain comparatively underutilized and face notable limitations~\cite{Martini2018}. Systematic TD tracking is still uncommon, and although many TD tools rely on techniques like static analysis (similarly to other quality tools)~\cite{Rios2018}, they often fall short of estimating the interest or likelihood associated with debt items, which hampers effective communication and prioritization \cite{Avgeriou2021}.

Practitioners also report practical hurdles when adopting TDM tools, including high error or false positive rates~\cite{Malakuti2021}, complex configuration~\cite{Biazotto2025}, limited alignment with existing workflows (e.g., integration to IDE or CI/CD embedding, and insufficient explainability \cite{Biazotto2025}. These limitations reduce both adoption and impact. Addressing them can enable \cite{Biazotto2023,Junior2022,Biazotto2025}: (i) more contextually aware and usable tools that cover a wider set of TD types; (ii) the combination of multiple information sources; (iii) actionable guidance, particularly around impact and consequences; and (iv) seamless integration and higher levels of automation throughout the modern toolchains.

Several studies have examined the usage of TDM tools in both industry and open-source projects~\cite{Khomyakov2019,Tornhill2022,Verdecchia2024}. Avgeriou et al. \cite{Avgeriou2021}, for instance, assessed tools to measure code, design, and architecture-related debt, analyzing characteristics, popularity, and available empirical evidence (including evaluations in industrial settings). While that work catalogs and characterizes the tools, our focus is on concerns surrounding their usage, specially in open-source software (OSS) context), bringing the practitioner's perspective.

Other literature \cite{Silva2022,Biazotto2023,Junior2022} shows that stakeholders often rely on isolated TDM tools. Junior et al.~\cite{Junior2022} highlighted the need for simpler integration of these tools into the prevailing development stack, moving toward a more holistic approach to TDM (i.e., supporting more activities with tools). Similarly, Biazotto et al.~\cite{Biazotto2023} argued for greater levels of automation to reduce the developer workload associated with TDM. While these contributions advance the theoretical understanding of tool usage, they offer limited practical guidance, an evidence gap that our study addresses. Our study tackles this gap by further investigating the usage of TDM tools in practice, helping both identify tools' limitations and provide guidance for developing management strategies (e.g., by following common patterns to add TDM tools to CI/CD pipelines).

\subsection{CI/CD and Configuration Anti-patterns}
\label{sec:rw-ci-cd}

CI/CD aims to make integrating and releasing changes safe, fast, and routine, providing rapid feedback through automated build, test, and deployment stages~\cite{FITZGERALD2017176}. In principle, such pipelines serve as quality gates that solve problems early and keep software in a releasable state. A CI/CD workflow (or pipeline) specifies an ordered and sequential set of stages that defines the process to build and deploy an application. Each stage has one or more jobs, which run in parallel. Such jobs have a set of phases (e.g., before\_install, script, after\_deploy), each one running a specific command\footnote{\url{https://docs.travis-ci.com/user/job-lifecycle/}}.

For example, consider Listing~\ref{lst:travis-python-example}, which shows a \travis file for a Python project.
This file defines a list of stages (1) to build and deploy the project. In this example, three stages were defined: lint (2), test (3), and deploy (4). Each stage has at least one job (e.g., 2.1); the ``test'' stage (3) has two jobs (i.e., 3.1 and 3.2). During execution, stages (2), (3) and (4) are executed sequentially, while jobs (3.1) and (3.2) are executed in parallel within the same stage. Since a stage can have one or many jobs, it can be a dedicated stage (i.e., with a single job, as stage 2, which runs only \tool{Flake8}) or mixed (i.e., many jobs, as stage (3)). The job itself can also be dedicated (i.e., executes only one action) or mixed (i.e., running multiple scripts). Although this structure provides significant flexibility for testing an application, pipelines themselves are software artifacts that must be designed and maintained;  when their configuration is degraded, the feedback becomes noisy, delayed, or misleading.

Prior work has cataloged recurring configuration anti-patterns (also called configuration smells or bad practices) that erode the potential benefits of CI/CD. Examples include overly long or monolithic builds, late integration/merging, silently ignored failures (e.g., \traviscommand{allow\_failures}), and broken or unstable main branches. Subsequent research proposed linters and analyzers to detect such problems. For instance, Vassallo et al.~\cite{Vassallo2019} introduced a reporting approach, CI-Odor, that automates the detection and longitudinal reporting of CI anti-patterns and pipeline decay, helping teams monitor when and where configurations start to drift. Complementary studies further show that pipelines can produce misleading outcomes (for example, by masking failures), compromising developer trust, and ultimately the value of automation~\cite{Gallaba2020}.

\begin{lstlisting}[style=yamlstyle,  basicstyle=\ttfamily\footnotesize, caption={Example of a \travis file for a Python project}, label={lst:travis-python-example}]
language: python
os: linux
stages: [lint, test, deploy] (1)
jobs:
  include:
    - stage: lint -- (2) --
      name: Lint -- (2.1) --
      install: pip install flake8
      script: flake8 src tests
    - stage: test -- (3) --
      name: Unit tests -- (3.1) --
      python: "3.11"
      install: pip install -r r.txt
      script: pytest -q
    - stage: test -- (3) --
      name: integration tests -- (3.2) --
      python: "3.11"
      install: pip install -r req.txt
      script: pytest -q -r
    - stage: deploy (4)
      if: tag IS present
      script: skip
      deploy: -- (4.1) --
        provider: pypi
        username: "__token__"
        password: $PYPI_TOKEN
\end{lstlisting}

Despite this promising work, we still lack consolidated, large-scale evidence on \textit{how} TDM tools are actually wired into real-world pipelines, \textit{which} configuration choices and anti-patterns co-occur with their usage, and \textit{what} actionable practices can help teams realize value consistently. The current study aims to address this gap.

\section{Study Design}
\label{sec:sd}

This section presents the objective and research questions (RQs) of this study, the data collection process, and how we analyzed the data to answer the RQs.

\subsection{Objective and Research Questions}
\label{sec:objective}
Considering the problems and gaps we presented in the two previous sections, the objective of this study, structured according to the Goal-Question-Metric template~\cite{vanSolingen2002}, is to %
\textit{``analyze \textbf{CI/CD configuration files} for the purpose of \textbf{identifying and classifying TDM tools} with respect to \textbf{their integration and configuration} from the point of view of \textbf{software developers} in the context of \textbf{open-source projects in GitHub}.''} To achieve this objective, we defined the following RQs:

\begin{itemize}
    \item \rqn{1} - \textbf{How are TDM tools integrated in CI/CD pipelines?} \\
    In this RQ, we investigate which TDM tools are commonly integrated into CI/CD pipelines, and how this integration usually occurs (i.e., by directly calling the tool or in an external script). Investigating the type of execution helps to understand how easily TDM logic can be maintained within pipelines (e.g., if TDM tools are in external scripts, more artifacts should be updated to change TDM logic). We also investigate if tools are usually used individually or co-occur with other tools. The answer to \rqn{1} can help practitioners evaluate their own tools and pipelines and be aware of alternatives for configurating those pipelines.
    
    \item \rqn{2} — \textbf{When are TDM tools integrated into CI/CD pipelines?}\\
    The stages, jobs, and phases at which a certain TDM tool is integrated into pipelines determine, among others, the visibility of the TDM feedback. For example, a phase indicates whether tools act as ``gates'' (i.e., the analysis happens before deployment and may block the build if there is any error) or as ``reporting'' (e.g., a linter that simply checks code quality, but does not affect the build process). In addition, the types of job/stage (e.g., dedicated or mixed) and their names help to organize TDM tools within the pipelines. Mapping tools across phases and job/stage structures can help to recommend strategies to integrate tools in TDM, increasing its value.

    \item \rqn{3} - \textbf{Which configuration anti-patterns are prevalent in pipelines with TDM tools?}\\
    Configuration anti-patterns (e.g., skipped failures) can hinder the value of TDM tools even when those tools are present. Quantifying the prevalence of such issues in pipelines that include TDM tools identifies the primary obstacles to trustworthy feedback. Together with \rqn{1} and \rqn{2}, \rqn{3} completes the progression from ``how'' TDM is integrated to ``when'' integration occurs in the pipelines and ``why'' it may succeed or fail in real-world CI/CD settings.
\end{itemize}

\subsection{Data Collection and Analysis}

In this section, we describe the methods we used and decisions we made during the data collection and analysis. All scripts used for these processes are reported in a replication package\rp. 

\vspace{.2cm}
\noindent\textbf{1) Selecting Context and CI/CD Managers to Investigate} 

We focus our investigation on GitHub, as it is one of the largest code repositories and has been extensively investigated in previous studies about TDM~\cite{KASHIWA2022, Tan2023, Tommasel2022}. Those studies provided evidence that GitHub provides the necessary variety of projects and teams to investigate TDM. As for CI/CD managers, we needed to select a manager that can be integrated with GitHub. While we had several options (e.g., Circle CI, GitHub Actions, and Jenkins), we decided to focus on Travis CI. Travis CI is one of the most popular CI/CD managers in GitHub~\cite{Rzig2022,Freitas2023}. Furthermore, since other CI services have a similar configuration syntax (YAML-based DSL), it is likely that our observations will be applicable to other CI services. In addition, previous work mentioned Travis CI as a good starting point for investigating tools in CI/CD pipelines~\cite{Chomatek2025}. Therefore, we argue that the focus on Travis CI and GitHub is sufficiently representative for a first exploratory study on TDM tools in CI/CD pipelines.

\vspace{.2cm}
\noindent\textbf{2) Retrieving Tavis CI Configuration Files}

To answer the proposed RQs, we need to identify GitHub projects that use Travis CI. As reported in previous studies~\cite{Vassallo2019, Gallaba2020, Chomatek2025}, the most efficient way to identify Travis CI is by searching for \travis files within the projects' root directories. To carry out this process, initially, we considered using the GitHub Search API\footnote{\url{https://docs.github.com/en/rest?apiVersion=2022-11-28}}, but that has significant limitations: first, it has a rate limit of 1,000 requests per hour, which deeply limits the feasibility to analyze a large amount of projects; second, as a HTTP-based API search, it imposes challenges for searching files with special characters (i.e., \travis). 

A second alternative is to use Google BigQuery\footnote{\url{https://cloud.google.com/bigquery?hl=pt_br}}, a serverless data warehouse on Google Cloud that lets one to run standard SQL queries over massive public datasets. Google BigQuery has a GitHub ``\textit{files}'' dataset (e.g., \textit{bigquery-public-data.github\_repos.files}), which is a snapshot of the files of public GitHub repositories stored in BigQuery. Using Google BigQuery, we retrieved a list of 526,864 repositories that have a Travis CI configuration file. 

One potential issue with the set of retrieved repositories from BigQuery is that we could not be sure that the GitHub snapshot in Google BigQuery was up-to-date. To mitigate this and expand the set of repositories, we used GH Archive\footnote{\url{https://www.gharchive.org}}, a repository that registers GitHub events (e.g., PushEvent and IssueCommentEvent). GH Archive provides hourly-based json files that contain all events registered during that period. In our study, we collected all repositories that had events between January 1st, 2025 and August 31st, 2025. We made a decision to focus on active projects instead of focusing on historical data, which might include discontinued features and/or practices. We deemed that this decision enabled us to investigate only updated projects, which have more potential to provide the current overview of CI/CD pipelines. 

The data collected from GH Archive include 6,652 json files, comprising around 430 GB of data. The json files contain around 1.1 billion events involving around 22 million unique GitHub repositories. We scanned each repository for \travis files within the project root directories, in the main branch (i.e., ``\textit{master}'' or ''\textit{main}''). In total, approximately 80,000 additional \travis were recovered for analysis. Therefore, around 600,000 Travis CI configuration files were recovered.

Considering that it is possible to use shell scripts\footnote{\url{https://docs.travis-ci.com/user/deployment/script}} to run TDM tools within Travis CI pipelines, we processed each \travis file to extract the shell scripts (e.g., \emph{analysis.sh}) that are executed in the CI/CD pipelines. The scripts must be stored within the repository, so we downloaded the scripts from GitHub. By doing this process, we collected around 50,000 scripts across the respositories that have a \travis file. This expands the corpus for analysis and provides a further understanding on how the tools are executed.

\vspace{.2cm}
\noindent\textbf{3) Identifying TDM tools}

Identifying CI/CD pipelines that encompass TDM tools is crucial for our study. While we acknowledge that any quality gate can be used as an alternative to mitigate TD~\cite{Arvanitou2019}, there is a variety of possible patterns to configure and execute such quality gates, making it infeasible to identify every pattern for our study. Hence, we decided to carry our study based on a sample of TDM tools that can be integrated with CI/CD.
To derive this sample of tools, we used a recent study on TDM \cite{Biazotto2023}, which reports on a set of 121 automation artifacts that can be used for TDM. From this set, we selected those tools which can be integrated with CI/CD pipelines as reported in \cite{Biazotto2023}. For each of those tools, we identified the patterns for using/executing them in CI/CD pipelines.  For example, SonarCloud can be executed within Travis CI by using the command ``\tool{sonar-scanner}''\footnote{\url{https://docs.travis-ci.com/user/sonarcloud/}}. Based on such patterns, we developed an analyzer that parses the Travis CI configuration files and searches for the patterns using regular expressions\footnote{\url{https://docs.python.org/3/library/re.html}}.

During the analysis of the list of tools, we noted some linters, such as pylint and ESlint\footnote{\url{https://eslint.org/}}. A linter is a static analysis tool that automatically checks the source code for errors, style issues, and suspicious patterns against defined rules. Considering that the list of linters reported in~\cite{Biazotto2023} is limited, and aiming to increase the representativeness of our list of tools,  we also identified patterns for running linters for the top-10 programming languages most used, according to the TIOBE raking\footnote{\url{https://www.tiobe.com/tiobe-index/}}. We are confident that our tool sample is representative for exploring TDM in CI/CD.

\vspace{.2cm}
\noindent\textbf{4) Identifying the Step that Executes the TDM tool}

The official documentation of Travis CI present an explanation of the jobs/stages of the CI/CD pipelines in Travis CI, as we described in Section~\ref{sec:rw-ci-cd}. Besides, previous studies~\cite{Gallaba2020, Chomatek2025} discuss such phases, and explain the goal of each one.  Inspired by both sources, we considered two characteristics of CI/CD steps to be investigated. First, the type of job/stage that runs a tool, i.e., dedicated stage, dedicated job within a stage (a job that runs only the TDM tool), and mixed job (a job that performs multiple actions). Second, regarding when the tool is executed, we consider if it is ran pre-deployment (i.e., as a quality gate) or after deployment (i.e., as a post-check to provide a quality report, but do not gate the build).

\begin{table*}[ht]
\footnotesize
\centering
\caption{CI/CD configuration anti-patterns (per Vassallo et al.) with selection and rationale for a TDM-in-CI/CD study}
\label{tab:ci-anti-patterns}
\begin{tabular}{p{0.067\textwidth} p{0.35\textwidth} p{0.02\textwidth} p{0.49\textwidth}}
\hline
\textbf{Anti-pattern} & \textbf{Definition (from Vassallo et al.)} & \textbf{Incl.} & \textbf{Rationale for inclusion/exclusion in this study} \\
\hline
Late Merging &
Agile teams develop in feature branches; integration effort and conflict potential increase if completed features are not integrated in a timely manner. &
Yes &
Directly affects timeliness of CI feedback and makes TD signals stale, reducing flow and team velocity; it is aligned with the need for integrating TDM with CI/CD and actionable flow metrics~\cite{Avgeriou2025}. Also matches Biazotto~\cite{Biazotto2025b}'s ``workflow consistency.'' \\

Skip Failed Tests &
Skipping a previously failing test ``fixes'' the build symptom without addressing the cause, undermining the safety of the test suite. &
Yes &
Hides debt and degrades trust in quality signals. Studying it supports the calls for reliable, visible data in CI/CD~\cite{Avgeriou2025} and Biazotto’s concerns about tool reliability and the need for human control over automation outputs~\cite{Biazotto2025b}. \\

Absent Feedback &
Developers miss required feedback if they are not automatically notified about relevant build events, especially failures.&
Yes &
A core visibility gap: without feedback, TD information does not reach practitioners. Biazotto~\cite{Biazotto2025b} shows many requirements are about communication (dashboards, IM, email) and notification configurability; \\

Email Notif. &
Email alone is an inappropriate single notification channel, as developers may lack access or notifications get lost among other messages. &
Yes &
Single-channel alerts exacerbate notification fatigue and are easy to miss. Biazotto reports preferences for configurable thresholds, summaries, and multiple channels~\cite{Biazotto2025b}; \\

Broken Release &
A broken build not fixed promptly prevents CI from assessing new changes. &
No &
Vassallo et al.~\cite{Vassallo2019} note teams are aware of this anti-pattern, thus reducing its risk to TDM. Besides, our study targets configuration choices that alter TD visibility and developer communication. \\

Slow Build &
Report increasing build times and outliers in build duration. &
No &
Our study focuses on patterns that directly shape TD feedback quality and communication. Performance concerns are acknowledged, but less central to TDM signal visibility than the selected four~\cite{Biazotto2025b}. \\

Aged Branches &
Infrequently synced branches diverge and become hard to integrate; warn when an open branch hasn’t been merged for a release. &
No &
Covered by \emph{Late Merging}; Vassallo merges the two due to overlap~\cite{Vassallo2019}. \\

Bloated Repo. &
Build artifacts/binaries should not be committed. &
No &
Repository hygiene is important but less directly tied to TD communication/integration in CI. Also dropped by Vassallo after survey, showing low relevance~\cite{Vassallo2019}. \\

Scheduled Builds &
Scheduled builds either rebuild changes unnecessarily or indicate changes are not automatically built. &
No &
Context-dependent and also dropped by Vassallo~\cite{Vassallo2019} due to low relevance. \\ \hline
\end{tabular}
\vspace{-20pt}
\end{table*}

\vspace{.2cm}
\noindent\textbf{5) Selecting and Identifying Configuration Anti-patterns}

Anti-patterns can have direct implications for TDM tools in CI pipelines. For example, if TDM tools are wired into stages that run infrequently (or post-merge only), the feedback loop becomes too long and opportunities for early remediation are lost. In the context of our study, we used the list of nine anti-patterns reported by Vassallo et al.~\cite{Vassallo2019}. Specifically, we considered four of those anti-patterns that can directly impact TDM and can be identified from the Travis CI files.  Table~\ref{tab:ci-anti-patterns} presents: (a) the definition of each selected anti-pattern; and (b) a rationale for including it in our study, considering its potential impact on TDM.
The selected anti-patterns can be identified considering only the \travis file, which also aligns with out data collection strategy. For each anti-pattern, we defined a set of rules to identify them in the CI/CD pipelines:

\begin{itemize}
    \item \textbf{Late Merging}: stage/job is limited to \traviscommand{type = push AND branch = main/master}.
        
    \item \textbf{Skip-on-Failures}: \traviscommand{jobs.allow\_failures} present
    
    \item \textbf{Absent Feedback}: No notification section, or no tokens for notification tools, neither the command \traviscommand{email} on the notifications section.
    \item \textbf{Email-Only Notifications}:  \traviscommand{notifications: { email: ... }} 
\end{itemize}

\vspace{.2cm}
\noindent\textbf{6) Analyzing Data}

To answer the RQs, we mainly relied on descriptive statistics. For \rqn{1}, we analyzed the share of adoption per tool, the type of execution (direct calls or script-based), as well as the co-integration of tools. To answer \rqn{2}, we analyzed the distribution of pre-deployment vs. post-deployment execution, the types of job (e.g., dedicated stage, dedicate jobs) as so as stage naming practices (i.e., the names of stages containing TDM tools). % considering the different tools. 
For \rqn{3}, we analyzed the prevalence of each anti-pattern, the co-occurrence of the anti-patterns, and the relation between tools and anti-patterns. We use frequency tables to carry out the analysis.

\section{Results}
\label{sec:results}

\subsection{How TDM tools integrate in CI/CD pipelines}
\label{sec:results-rq1}

To answer \rqn{1}, we analyze (a) which TDM tools have been integrated, (b) whether the CI/CD pipelines use direct or script-based invocations, and (c) whether multiple tools are integrated within the same pipeline. Regarding TDM tools, we identified a total of 38 tools integrated into CI/CD pipelines (Table~\ref{tab:tdm-tools-invoke}). Among them, \tool{Shellchek} is the most prevalent, appearing in 727 repositories, followed by \tool{Flake8} (724), \tool{Cppcheck} (332), \tool{Pylint} (315), and \tool{Govet} (292). Most of the identified tools are \emph{linters} or \emph{static analyzers} that primarily support the \emph{identification} of TD, indicating that TD detection is the dominant practice in CI/CD-based TDM. In contrast, fewer tools focus on \emph{measurement} (e.g., \tool{SonarQube}, \tool{Lattix}) or \emph{prevention} activities (e.g., \tool{Black}, \tool{Clang\_format}). Concerning TD types, the vast majority of tools address \emph{code debt}, followed by few tools targeting  \emph{build debt}. This distribution is expected, as source code is the primary artifact analyzed by automated CI/CD pipelines.

\begin{table}
\centering
\footnotesize
\caption{TDM tools in CI/CD pipelines}
\label{tab:tdm-tools-invoke}
\renewcommand{\arraystretch}{0.95}
\setlength{\tabcolsep}{3pt}
\begin{tabular}{p{1.4cm} p{1.8cm} p{1.7cm} p{1.3cm} p{0.4cm} p{0.4cm} p{0.4cm}}
\hline
\textbf{Tool} & \textbf{Tool type} & \textbf{TDM Activity} & \textbf{Debt Type} & \textbf{DC} & \textbf{SC} & \textbf{Pip.} \\
\hline
Shellcheck & Linter & Identification & Build & 69 & 672 & 727 \\
Flake8 & Linter & Identification & Code & 310 & 422 & 724 \\
Cppcheck & Static analyzer & Identification & Code & 129 & 205 & 332 \\
Pylint & Linter & Identification & Code & 89 & 226 & 315 \\
Govet & Static analyzer & Identification & Code & 47 & 245 & 292 \\
Clang\_format & Formatter & Prevention & Code & 78 & 213 & 272 \\
Eslint & Linter & Identification & Code & 66 & 175 & 241 \\
Phpcs & Linter & Identification & Code & 98 & 129 & 225 \\
Black & Formatter & Prevention & Code & 71 & 117 & 180 \\
Sonarqube & Static analyzer & Ident./Measur. & Code & 54 & 124 & 178 \\
Checkstyle & Linter & Identification & Code & 39 & 75 & 114 \\
Rubocop & Linter & Identification & Code & 70 & 44 & 110 \\
Golangci\_lint & Linter & Identification & Code & 71 & 28 & 97 \\
Clang\_tidy & Linter/Analyzer & Identification & Code & 31 & 64 & 94 \\
Phpstan & Static analyzer & Identification & Code & 72 & 5 & 76 \\
Cpplint & Linter & Identification & Code & 9 & 64 & 73 \\
Pmd & Static analyzer & Identification & Code & 17 & 54 & 71 \\
Sonarcloud & Static analyzer & Ident./Measur. & Code & 8 & 59 & 67 \\
Mypy & Static analyzer & Identification & Code & 27 & 36 & 62 \\
Tslint & Linter & Identification & Code & 54 & 5 & 59 \\
Coverity & Static analyzer & Identification & Code & 38 & 47 & 57 \\
Swiftlint & Linter & Identification & Code & 18 & 40 & 56 \\
Ruff & Linter & Identification & Code & 0 & 52 & 52 \\
Spotbugs & Static analyzer & Identification & Code & 19 & 26 & 45 \\
Yamllint & Linter & Identification & Build & 40 & 8 & 44 \\
Phpmd & Static analyzer & Identification & Code & 8 & 31 & 39 \\
Prettier & Formatter & Prevention & Code & 25 & 12 & 32 \\
Bandit & Static analyzer & Identification & Security & 15 & 11 & 25 \\
Lattix & Arch. analyzer & Measurement & Arch. & 0 & 24 & 24 \\
Findbugs & Static analyzer & Identification & Code & 1 & 20 & 21 \\
Staticcheck & Static analyzer & Identification & Code & 3 & 16 & 19 \\
Stylelint & Linter & Identification & Code & 10 & 1 & 11 \\
Psalm & Static analyzer & Identification & Code & 8 & 3 & 10 \\
Hadolint & Linter & Identification & Build & 5 & 5 & 9 \\
Detekt & Linter/Analyzer & Identification & Code & 6 & 1 & 7 \\
Swiftformat & Formatter & Prevention & Code & 0 & 5 & 5 \\
Brakeman & Static analyzer & Identification & Code & 1 & 3 & 4 \\
Ktlint & Linter & Identification & Code & 1 & 1 & 2 \\
\hline
\end{tabular}
\vspace{-17pt}
\end{table}

Considering the type of invocation of each tool, 2,466/3,684 pipelines invoke tools using external scripts (66.9\%), while 1,127/3,684 do so by calling the tool directly (30.6\%),  and 91/3,684 use both strategies (2.5\%). Therefore, maintainers often ``glue'' TDM tools through shell/auxiliary scripts rather than declaring them inline in Travis CI. This approach decentralizes logic, i.e., practitioners have CI/CD logic in both the configuration file and the scripts. Therefore, using external scripts can hide TDM checks from the top-level pipeline view. This can be a problem when the TDM logic needs to be maintained: multiple execution points (e.g., multiple scripts) make it harder to change tool configuration and increase maintenance costs, which might decrease the perceived value of TDM.

At the tool level, the highest volumes of direct calls (DC in Table~\ref{tab:tdm-tools-invoke}), are in \tool{Flake8} (310) and \tool{Cppcheck} (129). As for script calls (SC in Table~\ref{tab:tdm-tools-invoke}), \tool{Shellcheck} (672), \tool{Flake8} (422), and \tool{Govet} (245) are the most prevalent. We note that 35/38 tools are invoked using both strategies, suggesting no practitioner preference. For instance, \tool{Tslint} tends to be executed via direct calls because the integration between Travis CI and the tool is straightforward, whereas \tool{Sonarqube} is mainly used in scripts, even though the integration of \tool{Sonarcloud} with Travis CI is relatively simple. In summary, while highly context dependent, practitioners tend to use TDM tools mostly in script files.

\begin{figure}
    \centering
    \includegraphics[width=0.4\linewidth]{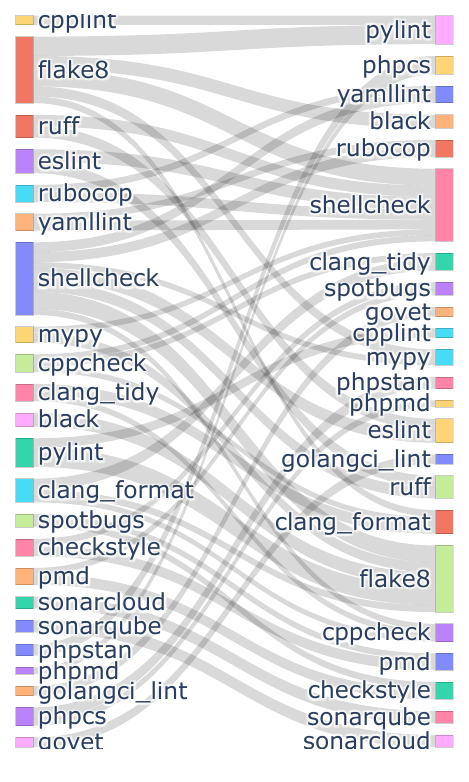}
    \caption{Tools that co-occur in at least 20 different pipelines}
    \label{fig:rq1-tools-co-ocurrence}
    \vspace{-35pt}
\end{figure}

Finally, we checked how many tools are typically integrated within the same pipeline. Overall, 2,900/3,684 (78.7\%) of repositories use exactly one TDM tool; 555/3,684 (15.1\%) use two; 174/3,684 (4.7\%) use three; and a 55/3,684 (1.5\%) use four or more. Therefore, most pipelines adopt a single primary tool rather than a broad toolchain. To extend this analysis, we examined stacks (co-occurrence of tools within the same repository) to understand common combinations of tools. Figure~\ref{fig:rq1-tools-co-ocurrence} depicts the tools that co-occur in at least 10 pipelines. While a full list of co-occurrences can be found in our replication package\rp, here we limit the analysis to pairs appearing in at least 10 pipelines to increase representativeness.

From the 38 identified tools, 36 were found to co-occur in pipelines. In total, we identified 142 different combinations, with \tool{Flake8}–\tool{Pylint} (77) and \tool{Flake8}–\tool{Shellcheck} (67) as the most common. Across ecosystems, multiple combinations support Python and Java projects (e.g., \tool{Flake8}–\tool{Pylint} and \tool{Pmd}–\tool{Checkstyle}). In addition, some co-occurrences focus on checking project infrastructure (e.g., \tool{Shellcheck}–\tool{Yamllint}), which points to a concern for keeping pipeline code consistent. Finally, we also noted cross-ecosystem co-occurrences, such as \tool{Flake8}-\tool{Cppcheck} (for Python and C/C++) and \tool{Eslint}–\tool{Pylint} (for JavaScript to Python). While stacks are not prevalent in our dataset, when teams do stack tools, they usually combine tools that are complementary within a single ecosystem (e.g., linting and bug-finding for Python) and, sometimes, pair the tools with configuration/scripting checks (e.g., shell checkers).

\begin{findingbox}
\footnotesize
\textbf{Answer to \rqn{1}:} TDM in CI/CD is carried out predominantly by linters, and \tool{Flake8}, \tool{Shellcheck}, \tool{Cppcheck}, \tool{Govet}, and \tool{Pylint} are the most integrated tools. Tools are mostly invoked through external scripts (67.7\%), indicating a preference for a separation of concerns between CI/CD steps and quality checks. Typical stacks pair a linter with a tool that provides deeper code analysis (e.g., code metrics) or a formatter (e.g., \tool{Phpstan} + \tool{Phpcs}), but most projects use just one tool.
\end{findingbox}

\vspace{1cm}
\subsection{When TDM Tools integrate in CI/CD pipelines}
\label{sec:results-rq2}

As we explained in Section~\ref{sec:rw}, a job is an isolated execution unit within a CI/CD pipeline (i.e., a self-contained set of steps), and jobs may be grouped into stages that run sequentially; a single pipeline may have multiple jobs and stages. To answer \rqn{2}, we analyze three aspects of jobs/stages:

\begin{enumerate}
    \item To understand whether practitioners separate TDM checks from other processes in pipelines, we classify jobs/stages into three types: dedicated stage, dedicated job, and mixed job. This helps us to clarify how the TDM tools are usually spread over the CI/CD stages and jobs.
    
    \item We also investigate whether TDM tools are run before the deployment, as a quality gate, or after the deployment, as a simple report. This helps us to understand the moment in which TDM processes are carried.
    
    \item Finally, we investigate stage naming. As we discussed in Section~\ref{sec:intro}, the pipelines themselves are artifacts that must be maintained over time. Explicitly defining stage names for TDM could simplify the maintenance of such pipelines, and it is important to report current practices for naming.
\end{enumerate}

We split the data set according to the type of execution, that is, \emph{directly} in \travis (3,283 jobs) or through a \emph{external script} (2,993 jobs), as we did in \rqn{1}. Since pipelines might have multiple jobs, the number of jobs (6,276 jobs) is higher than the number of pipelines (3,684 pipelines) 

First, we identified the types of jobs/stages: (i) \textbf{dedicated stage}, meaning the whole stage (i.e., group of jobs) focuses on configuring and running the TDM tool; (ii) \textbf{dedicated job within stage}, which means that a job within the stage focuses on running the TDM tool; and (iii) \textbf{mixed job}, where the TDM tool shares the main job phases with other tasks (e.g., tests / build).

\begin{figure}[h]
    \centering
    \includegraphics[width=0.65\linewidth]{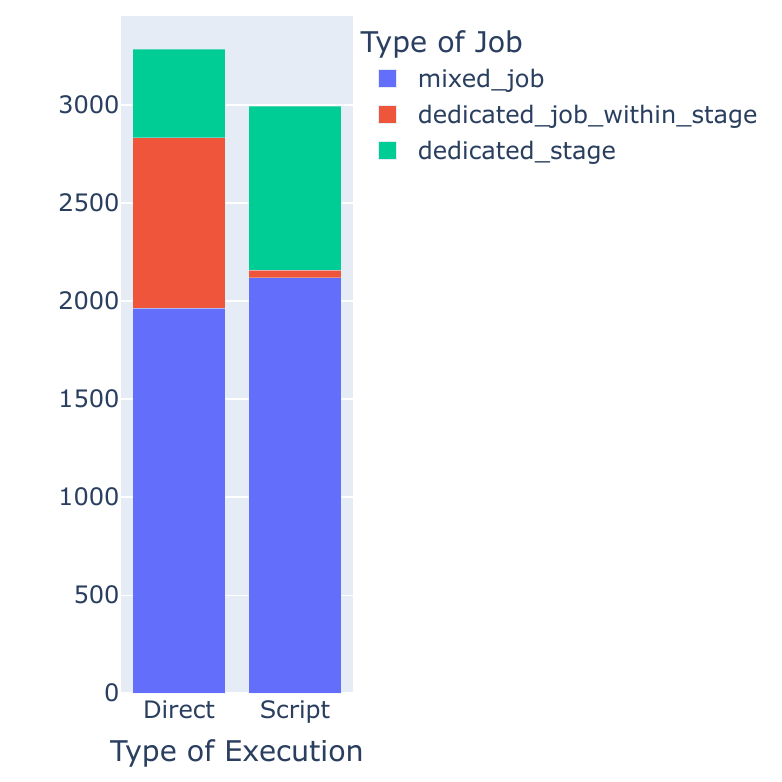}
    \caption{Types of stage/job}
    \label{fig:rq2-job-type}
    \vspace{-20pt}
\end{figure}

Figure~\ref{fig:rq2-job-type} reports the distribution of types for script-based and direct executions. In direct calls, \emph{mixed job} pipelines account for a little more than a half of the cases (1,963/3,283), while script-based setups raise this to about two-thirds (2,119/2,993). Direct calls show a larger portion of \emph{dedicated job within stage} (870/3,283), whereas \emph{dedicated stage} (450/3,283) is a minority. In script-based TDM, dedicated stages (836/2,993) are more common than in direct calls.

Regarding the moment in which TDM tools are executed, pre-deployment executions are most prevalent in both direct calls (3,261/3,283) and script calls (2,907/2,993). This suggests that most TDM tools are used as quality gates. Optionally, some tools are also used for \emph{reporting} after deployment. Such tools include \tool{Coverity} and \tool{Flake8}, indicating that tools providing reports that are not critical (e.g., test coverage) can be run later in the pipeline. The few post-deployment executions are mainly present in script-based setups. Since scripts are more flexible and enable combing runtime information (e.g., deployment duration) with other static checks, practitioners might prefer this type of execution for post-checks.
 
Regarding stage naming, most TDM executions live in an unnamed stage/job (5,057). This is largely a consequence of Travis CI’s defaults: when jobs are not explicitly grouped into stages, the pipeline runs in a single \emph{implicit} stage, named \traviscommand{test}\footnote{We decided to use ``implicit'' instead of ``test'' to avoid wrongly capturing named stages}. In practice, the lack of stage names mean that many pipelines operate as a single, bulk CI stage. This explains why the vast majority of direct executions fall into unnamed stages, since practitioners are not forced to name stages, so the default behavior silently absorbs TDM into the implicit \traviscommand{test} stage, making configurations simpler to authors, but also easier to ignore and harder to review at a glance.

When teams name stages explicitly, the labels usually indicate a quality-related concern (e.g., \emph{Lint/lint} and \emph{Code Quality}). Thus, when explicit, TDM tools appear to be assigned to stages more specific and appropriate, instead of more general (e.g., test) or mixed ones. Table~\ref{tab:stages-naming} shows a list of stage names that appear in at least five pipelines. A complete list of names is available in our replication package.

\begin{table}[h]
\small
\centering
\caption{Number of jobs per stage name (at least 20 stages)}
\begin{tabular}{p{4.5cm} r c c}
\hline
\textbf{Stage Name} & \textbf{Direct} & \textbf{Scripts} & \textbf{Total}\\ \hline
implicit & 2442 & 2615 & 5057 \\
lint & 105 & 39 & 144 \\
test & 90 & 0 & 90 \\
Lint & 63 & 0 & 63 \\
macOS Build & 0 & 61 & 61 \\
Code Quality & 52 & 0 & 52 \\
build & 33 & 3 & 36 \\
Coverity & 29 & 0 & 29 \\
Build, Test \& Check Code Quality & 27 & 0 & 27 \\
Coding standard & 21 & 0 & 21 \\
verify & 21 & 0 & 21 \\
\hline
\end{tabular}
\label{tab:stages-naming}
\vspace{-7pt}
\end{table}

\begin{findingbox}
\textbf{Answer to \rqn{2}} Teams frequently place TDM tools to execute checks before the deployment. Moreover, most jobs are mixed, and TDM tools execute alongside with other tools or processes (e.g., pipeline configurations). In addition, most stages names are implicit; for explicit names, ``\textit{lint}'' and ``code quality'' are the most common. 
\end{findingbox}

\subsection{CI/CD configuration anti-patterns in TDM}
\label{sec:results-rq3}

Across 3,684 pipelines, \emph{Absent Feedback} is the dominant anti-pattern (2,493 or 67.7\%), while \emph{Skip-on-Failure} (565 or 15.3\%) and \emph{Late Merging} (412 or 11.2\%) are less frequent. \emph{Email-only} notifications appear in 251 pipelines (6.8\%). The distribution shows that most pipelines exhibit exactly one anti-pattern (2,482 or 65.7\%), with fewer showing two (649 or 17.2\%) and only a small fraction exhibiting all three (16 or 0.5\%); 628 pipelines (16.6\%) show none.

We also identified some co-occurrences between the anti-patterns, as presented in Table~\ref{tab:rq3-co-occurrence}. \emph{Late Merging} co-occurs with \emph{Absent Feedback} in 226 cases (54.9\% of \emph{Late Merging} pipelines), and \emph{Skip-on-Failure} co-occurs with \emph{Absent Feedback} in 371 cases (65.7\% of \emph{Skip-on-Failure} pipelines). \emph{Late Merging} and \emph{Skip-on-Failure} rarely occur together (23 pipelines). Since the \textit{Absent Feedback} is the most common anti-pattern, it was expected that it would co-occur with other anti-patterns more frequently. 

\begin{table}[h]
\centering
\caption{Co-occurrences of anti-patterns}
\begin{tabular}{p{3.3cm}p{0.9cm}p{0.9cm}p{0.9cm}p{0.8cm}}
\toprule
 & \textbf{Absent Feedback} & \textbf{Skip-on-Failure} & \textbf{Late Merging} & \textbf{Email-only} \\
\midrule
Absent Feedack & - & 371 & 226 & 0 \\
Skip-on-Failure & 371 & - & 23 & 29 \\
Late Merging & 226 & 23 & - & 20 \\
Email-only Notifications & 0 & 29 & 20 & - \\
\bottomrule
\label{tab:rq3-co-occurrence}
\end{tabular}
\vspace{-13pt}
\end{table}

We also examine tools that are frequently related to anti-patterns. For this, we use Tables~\ref{tab:AF}, \ref{tab:LM}, \ref{tab:SoF}, and \ref{tab:EoN} that show the number of pipelines each tool is involved in, the number of pipelines with a certain anti-pattern for that tool, and the percentage of pipelines with that anti-pattern. For instance, in Table~\ref{tab:AF}, we show that \textit{Absent Feedback} occurs in 94.9\% of the pipelines with \tool{Tslint}.  

\begin{table}[h]
\centering
\caption{Absent Feedback (AF) among projects using each tool (with AF $\geq 50$).}
\label{tab:AF}
\begin{tabular}{lp{1.2cm}p{1.2cm}p{1.2cm}p{1.2cm}}
\toprule
\textbf{Tool} & \textbf{Pipelines} & \textbf{with AF} & \textbf{\% AF} \\
\midrule
Tslint & 59 & 56 & 94.9\% \\
Clang\_format & 272 & 239 & 87.9\% \\
Golangci\_lint & 97 & 85 & 87.6\% \\
Clang\_tidy & 94 & 81 & 86.2\% \\
Black & 180 & 151 & 83.9\% \\
Sonarcloud & 67 & 53 & 79.1\% \\
Govet & 292 & 230 & 78.8\% \\
Phpstan & 76 & 59 & 77.6\% \\
Eslint & 241 & 179 & 74.3\% \\
Flake8 & 724 & 517 & 71.4\% \\
Sonarqube & 178 & 124 & 69.7\% \\
Checkstyle & 114 & 79 & 69.3\% \\
Phpcs & 225 & 141 & 62.7\% \\
Cppcheck & 332 & 199 & 59.9\% \\
Pylint & 315 & 182 & 57.8\% \\
Shellcheck & 727 & 407 & 56.0\% \\
Rubocop & 110 & 56 & 50.9\% \\
\bottomrule
\end{tabular}
\end{table}

\emph{Absent Feedback} is widespread among the pipelines, as Table~\ref{tab:AF} shows. The tools, such as \tool{Golangci\_lint} (87.9\%) and TSLint (94.9\%), are highly correlated with this pattern. This might indicate that simpler checks, such as evaluating shell structures, do not impose strong concerns on developers. Therefore, notifications may be ignored, because immediate actions are not necessary.

As listed in Table~\ref{tab:LM}, \tool{Black} (38.3\%) and \tool{Pmd} (28.2\%) are likely to appear in pipelines with \textit{Late Merging}. On the other hand, \tool{Pylint} is much less frequent (4.4\%). These higher rates among static code analyzers (e.g., \tool{Pmd}) indicate that more complex analysis (e.g., identification of code smells with \tool{Checkstyle}) are often performed only when the code is merged to the main branch, which might not be ideal and could potentially delay TDM.

\begin{table}
\centering
\caption{Late Merging (LM) among projects using each tool (with LM $\geq 10$).}
\label{tab:LM}
\begin{tabular}{lp{1.2cm}p{1.2cm}p{1.2cm}p{1.2cm}}
\toprule
\textbf{Tool} & \textbf{Pipelines} & \textbf{with LM} & \textbf{\% LM} \\
\midrule
Black & 180 & 69 & 38.3\% \\
Pmd & 71 & 20 & 28.2\% \\
Checkstyle & 114 & 32 & 28.1\% \\
Sonarcloud & 67 & 16 & 23.9\% \\
Shellcheck & 727 & 172 & 23.7\% \\
Spotbugs & 45 & 10 & 22.2\% \\
Sonarqube & 178 & 25 & 14.0\% \\
Rubocop & 110 & 11 & 10.0\% \\
Eslint & 241 & 16 & 6.6\% \\
Clang\_format & 272 & 18 & 6.6\% \\
Govet & 292 & 15 & 5.1\% \\
Phpcs & 225 & 11 & 4.9\% \\
Flake8 & 724 & 34 & 4.7\% \\
Pylint & 315 & 14 & 4.4\% \\
\bottomrule
\end{tabular}
\vspace{-10pt}
\end{table}

\begin{table}[h]
\centering
\caption{Skip-on-Failure (SoF) among projects using each tool (with SoF $\geq 10$).}
\label{tab:SoF}
\begin{tabular}{lp{1.2cm}p{1.2cm}p{1.2cm}p{1.2cm}}
\toprule
\textbf{Tool} & \textbf{Pipelines} & \textbf{with SoF} & \textbf{\% SoF} \\
\midrule
Coverity & 57 & 42 & 73.7\% \\
Phpstan & 76 & 53 & 69.7\% \\
Lattix & 24 & 14 & 58.3\% \\
Phpcs & 225 & 97 & 43.1\% \\
Clang\_tidy & 94 & 39 & 41.5\% \\
Cpplint & 73 & 16 & 21.9\% \\
Cppcheck & 332 & 68 & 20.5\% \\
Clang\_format & 272 & 52 & 19.1\% \\
Flake8 & 724 & 133 & 18.4\% \\
Shellcheck & 727 & 61 & 8.4\% \\
Pylint & 315 & 26 & 8.3\% \\
Govet & 292 & 23 & 7.9\% \\
Sonarqube & 178 & 10 & 5.6\% \\
Eslint & 241 & 13 & 5.4\% \\
\bottomrule
\end{tabular}
\vspace{-5pt}
\end{table}

For \emph{Skip-on-Failure}, practices are mixed, as Table~\ref{tab:SoF} shows. Tools such as \tool{PHPStan} (69.7\%) and
\tool{Coverity} (73.7\%) are often configured to allow failures. Other tools, such as \tool{Eslint} (5.4\%) and \tool{Sonarqube} (5.6\%), usually block the execution in case of failures. This pattern indicates that tools focused on code style and formatting (e.g., \tool{Flake8}) or test coverage (e.g., \tool{Coverity}) are more frequently treated as non-blocking checks and do not enforce a pipeline to stop in case of violations. On the other hand, linters and static analyzers (e.g., \tool{Eslint}), which provide more detailed reports on code quality (e.g., violations), are often configured to enforce the pipeline to stop in case of failures.

For \emph{Email-only Notifications}, Table~\ref{tab:EoN} shows that tools such as \tool{Cpplint} (47.9\%) and \tool{Spotbugs} (31.1\%)\ often rely on email as notification channel; \tool{Shellcheck} (2.7\%) and \tool{Cppcheck} (3.0\%) usually uses other channels, such as Slack. This pattern suggests that pipelines running tools with more detailed reports (e.g., the smells identified by \tool{Spotbugs}) usually send results by email. 

\begin{table}[h]
\centering
\caption{Email-only Notifications (EoN) among projects using each tool (with EoN $\geq 10$).}
\label{tab:EoN}
\begin{tabular}{lp{1.2cm}p{1.2cm}p{1.2cm}p{1.2cm}}
\toprule
\textbf{Tool} & \textbf{Pipelines} & \textbf{with EoN} & \textbf{\% EoN} \\
\midrule
Cpplint & 73 & 35 & 47.9\% \\
Spotbugs & 45 & 14 & 31.1\% \\
Pmd & 71 & 16 & 22.5\% \\
Pylint & 315 & 67 & 21.3\% \\
Phpcs & 225 & 47 & 20.9\% \\
Checkstyle & 114 & 19 & 16.7\% \\
Flake8 & 724 & 47 & 6.5\% \\
Govet & 292 & 18 & 6.2\% \\
Clang\_format & 272 & 13 & 4.8\% \\
Cppcheck & 332 & 10 & 3.0\% \\
Shellcheck & 727 & 20 & 2.8\% \\
\bottomrule
\end{tabular}

\end{table}

\begin{findingbox}
\textbf{Answer to \rqn{3}:} Across 3,684 pipelines, \emph{Absent Feedback} is the most common anti-pattern. Regarding the tools in pipelines with anti-patterns, linters and static code analyzers are less often related to \emph{Late Merging} and \emph{Skip-on-Failure} (i.e., such analysis runs earlier in the pipeline and enforces the build stop in case of failures).
\end{findingbox}

\section{Discussion}
\label{sec:discussion}
\subsection{Interpretation of the results of \rqn{1}}
\label{sec:discussion-rq1}
The results of \rqn{1} indicate that TD identification is the most prevalent activity in CI/CD pipelines, since most tools are linters/static analyzers (e.g., \tool{Flake8}). These results are consistent with other recent literature~\cite{Silva2022, Junior2022, Biazotto2025b}, which states that identification is the most performed TDM activity. Although adding TD identification tools helps to improve awareness of TD, the potential for performing other TDM activities (e.g., TD repayment) is underexplored. Thus, we advise researchers to investigate how TD measurement (quality gates, trend tracking), prevention (auto-fixers, formatters), and repayment (e.g., refactoring suggestions) activities could be inserted into CI/CD pipelines.

We also found that most pipelines run single TDM tools, while few TDM stacks, i.e., combinations of TDM tools, also appear. Considering the flexibility provided by CI/CD pipelines, it would be most beneficial if stacks focusing on multiple TDM activities were implemented. We thus  recommend that researchers investigate defining, and testing TDM tool stacks. As for practitioners, a suggestion is to start integrating TDM into CI/CD with a static code analyzer and potentially combining other tools (such as a linter) for the same ecosystem (e.g., JavaScript). This would help to increase TD visibility without disrupting existing pipelines.

Regarding tool execution, most pipelines invoke tools via external scripts. This practice may help reusing commands across jobs and projects; it can also help scale TDM while reducing the effort to maintain TDM tools. In addition, external scripts might help keep the Travis CI YAML concise, promoting a more coherent separation of concerns. However, moving TDM logic to scripts reduces the visibility of quality, making reviews and quick comprehension harder, especially for newcomers. This trade-off may be more critical for OSS, because the number of (new) contributors is usually higher in OSS than in industry. Nevertheless, we advise all practitioners to pay attention to stage naming when using external scripts and to use mechanisms like log messages or notifications, to make the TD checks more visible.

\subsection{Interpretation of the results ot \rqn{2}}
\label{sec:discussion-rq2}
The results from \rqn{2} show that most TDM tools run in pre-deployment phases, which is consistent with common CI/CD practice around quality gates. As discussed in Section~\ref{sec:sd}, any quality gate can support TDM~\cite{Arvanitou2019}, and our data corroborate this because most tools are \emph{actually} executed before deployment (i.e., as a gate). As an implication, practitioners could increase the value of these gates by adopting TDM tools that further explain the bad effects of TD (e.g., interest or maintainability indexes). Practitioners could also combine pre-deployment checks, which make TD visible earlier, with simple post-deployment reports (e.g., test coverage), which provide additional information to support planning TDM. Finally, researchers can investigate whether using tools as gates vs. for reporting, impacts the perceived usefulness and value of TDM tools.

Most TDM tools are present in mixed jobs, which once again indicates that TDM is treated as any other quality check within the pipeline. Furthermore, when TDM logic is pushed into external scripts, it might be mixed with other non quality-related tasks, such as environment preparation and caching. This somehow reduces the coherence of the TDM checks and might reduce the perception of its value. In direct calls, there is a relatively higher share of dedicated jobs within a stage. This could happen simply because, with seamless integration between the TDM tool and the pipeline (i.e., a single command), dedicated stages become simpler to maintain and make the configuration file more consistent. Based on these observations, we deem that practitioners should avoid mixing TDM with non-quality-related processes (e.g., pipeline configuration), ideally keeping TDM checks as a single stage in the pipeline. This would provide both more flexibility (e.g., multiple TDM checks can be combined) and more coherence, simplifying the maintenance of TDM tools and pipelines

Finally, naming practices show that many tool executions occur in unnamed /implicit stages. In practice, this treats TDM tools as ``just another step.'' Similar to the problems identified in \rqn{1}, unnamed stages might hinder the understandability of YAML files and keep TDM hidden from maintainers. In contrast, when developers name stages, the labels are related to quality (e.g., \emph{Lint}, \emph{Code Quality}, \emph{Static Analysis}). However, we did not identify any specific stage name for TD (e.g., debt or tech debt). A good practice for practitioners would be to clarify the stage names and clearly define when TDM is being carried out.

\subsection{Interpretation of the results to \rqn{3}}
\label{sec:discussion-rq3}
The prevalence of \emph{Absent Feedback} (2,493/3,684) suggests that many teams successfully run TDM tools but fail to communicate their results. In practice, this reduces the effectiveness of the feedback and hinders awareness of TD. Hence, the support provided by static analysis and quality checks has not been exploited at its maximal potential. Researchers should investigate how to measure and improve the observability of TD-related failures in CI/CD pipelines. A potential direction is to define and track visibility metrics and evaluate their relationship with tool adoption and code review effectiveness (i.e., whether increased visibility leads to more effective remediation actions for TD).

Regarding \emph{Late Merging} (412/3,684), a core problem is that it shifts attention from prevention (that is, managing TD in pull requests) to detection after integration. The results also indicated that static analyzers (e.g., \tool{Pmd}) are mostly executed in main branches only, while tools with less complex analysis (e.g., \tool{Pylint}) tend to run in every branch. Overall, considering that \emph{Late Merging} appears in less than 10\% of all pipelines, we can infer that practitioners are already using TDM tools as gates (i.e., checking TD before merging the code). 
Nonetheless, the higher rates on static analyzers, such as \tool{Checkstyle}, indicate that deeper TD analysis (e.g., smells detection) are being carried only when the code is merged to the main branch. Therefore, we caution practitioners that such practices might be risky and should be avoided. For researchers, they could investigate the combinations of pre- and post-merge checks. This would both raise the awareness of practitioners about the problems of \emph{Late Merging} and define scenarios in which post-merge checks could be safely used. This understanding might lead to the definition of guidelines or even CI/CD templates focused on TDM; those templates could be easily replicated in different projects.

Finally, \emph{Skip-on-Failure} (565/3,684) hinders the ``\textit{gatekeeping}'' aspect of TD tools in CI/CD pipelines. This can explain how pipelines can sometimes appear healthy while accumulating unmanaged debt. According to the results, linters appear to be less critical for practitioners and more prone to this anti-pattern. Specifically for TDM, we advise practitioners to avoid skipping failures, since this can hide the TD. Another suggestion for practitioners is to have separated jobs for higher-risk checks (e.g., architectural issues with \tool{Checkstyle}) and avoid skipping failures on those. If necessary, allowing-failure policies must be rare, deliberate, and documented to ensure the reliability of the pipeline. 

\section{Threats to Validity}
\label{sec:tov}
\textbf{Construct validity:} Our identification of Travis CI configuration files relies on BigQuery and GitHub Archive, and we may have missed projects (e.g., configuration files in non-default branches) or included false positives (e.g., non-Travis CI YAMLs named as \travis). To mitigate this threat, we developed scripts that check whether the minimal configurations of Travis CI (e.g., a script section) are present and removed the files that did not have such configurations. The operationalization for identifying CI/CD anti-patterns (e.g., \emph{Late Merging}, \emph{Skip-on-Failure}) may over- or under-flag cases; we derived characteristics of each anti-pattern from published definitions (e.g., \cite{Vassallo2019}) and defined the identification rules considering the structure of Travis CI files (as we describe in Section~\ref{sec:sd}). However, Travis's configuration files are highly heterogeneous, and it is not possible to ensure that we were able to capture all possible types. As a mitigation action, the scripts we used to process the Travis CI files are based on the official documentation only\footnote{\url{https://docs.travis-ci.com/user/job-lifecycle/}}.

\vspace{0.2cm}
\textbf{External validity:} As our dataset focuses on GitHub repositories and Travis CI, we cannot generalize the results to other platforms (e.g., GitLab, Bitbucket) or CI/CD platforms (e.g., GitHub Actions, GitLab CI, CircleCI), although other CI/CD platforms are mostly similar to Travis CI (e.g., centered around the concept of jobs or configured using YAML files). Hence, replications are needed to increase external validity. Similarly, the set of tools we analyzed is limited. To mitigate this threat, we considered not only the set of tools identified in a previous study \cite{Biazotto2023} but also the list of other linters (i.e., linters for the top-10 most used programming languages, according to the TIOBE raking\footnote{\url{https://www.tiobe.com/tiobe-index/}}). While we acknowledge that the investigated tools are sufficiently representative for an initial exploratory study, replications are needed to improve the generalizability of our results.

\vspace{0.2cm}
\textbf{Reliability:} To mitigate threats to the reliability of the study, we describe the data acquisition process in as much detail as possible. More importantly, the data set curated through our investigation and the scripts used for data collection and analysis are publicly available\rp. 

\section{Conclusion and Future Work}
\label{sec:conclusion}
The main finding of this study is that CI/CD-based TDM is mainly destined to detect TD, rather than communicate it or make it visible. Our results show that linters are the dominant tools, and nearly two-thirds of CI/CD pipelines invoke TDM tools via scripts. We see two research directions from our results: (i) cross-platform (e.g., using Gitlab or Github Actions) and longitudinal studies are essential to measure the benefits of TDM in CI/CD; and (ii) while integrating TDM tools using external or direct calls might be help the separation of concerns between pipelines and TDM check, it is still necessary to survey practitioners to further understand the rationale for such decisions. Both research directions can enable the definition of guidelines for adopting TD tools in continuous and agile environments.

\bibliographystyle{IEEEtran}
\bibliography{references}

\end{document}